\documentstyle[aps,preprint]{revtex}
\begin{document}
\draft
\title{Macroscopic and mesoscopic matter waves}
\author{Ram K. Varma$^{\dag}$ }
\address{Physical Research Laboratory, Ahmedabad 380 009, INDIA}
\maketitle
\begin{abstract}
\vspace*{0.2in}

It has been shown earlier [3,4,6]
that matter waves which are known to
lie
typically in the range of a few {\AA}ngstrom, can also manifest in the
macrodomain with a wave length of a few centimeters, for electrons
propagating along a magnetic field. This followed from the predictions of
a probability amplitude theory by the author[1,2] in the classical
macrodomain of the dynamics of charged particles in a magnetic field. It
is shown in this paper
that this case
constitutes only a special case of a generic situation whereby composite
systems such as atoms and molecules in their highly excited internal
states,can exhibit matter wave manifestation in macro and mesodomains.
The wave length of these waves is determined, not by the mass of the
particle as in the case of the de Broglie wave, but by the frequency
$\omega$, associated with the internal state of excitation,  and is given
by a nonquantal expression, $\lambda =2\pi v/\omega$, $v$ being the
velocity
of the particle. For the
electrons in a
magnetic field the frequency corresponds to the gyrofrequency, $\Omega$
and
the nonquantal wave
length is given by $\lambda = 2\pi v_{\parallel}/\Omega$; $v_{\parallel}$
being the velocity of electrons along the magnetic field.\\\

\dag e-mail rkvarma@prl.ernet.in

\end{abstract}
\newpage

{\bf 1. Introduction}

The de Broglie matter waves associated with quantum particles have a
wavelength typically a few {\AA}ngstroms ($\lambda =
\hbar/mv$) essentially
because of the small value of $\hbar$. The question may be asked however,
whether matter can exhibit its wave aspect in the macrodimensions as well,
not in the sense of macroscopic correlated quantum systems such as
superfluids or superconductors, but in the manner and spirit of de Broglie
waves associated with single particles. Following the development of the
concept of macroscopic matter waves through the theory of Ref.[1],
reinforced by a more recent work by the author [2], we had
demonstrated the existence of
such a wave behavior for electrons propagating along a magnetic field
having a wavelength independent of $\hbar$ and typically in the range of a
few centimeters [3,4]. We wish to show here that such a wave
manifestation
is not
entirely peculiar to this system but is a generic property of composite
bound systems in their highly excited internal states approaching the
classical limit. The wavelength of these new matter waves is related, not
to the masses of these particles as in the case of de Broglie waves, but
to the frequency associated with their internal state of excitation. This
is an entirely new wave manifestation of matter not hitherto pointed out.
We also predict the existence of such macroscopic or mesoscopic waves with
atoms and molecules.

We first present here the development of the concept of macroscopic matter
waves in the context of charged particle dynamics in a magnetic field,
as expounded in the theories of Ref.[1,2] which is not widely known. After
having captured the essence of such a
behaviour for this system we shall show later how one could extend 
this concept to other systems, essentially composite systems such
as atoms and molecules.

Even though the concept of the macroscopic matter waves in relation to the
charged particle dynamics along a magnetic field actually followed from
the
theories of Ref[1,2], we present here first a direct quantum mechanical
derivation of the macroscopic form of the wave function which is
consistent with the form obtained from Ref[1,2] and which can account for
the rather astonishing observations reported [3,4] in this connection,
which were predicted by the theory. We
later extend these considerations to other systems such as atoms and
molecules, and in fact any composite system.

{\bf 2. Macroscopic wave function and matter waves for  charged
particles in magnetic field}

A charged particle in a magnetic field in the classical mechanical domain
corresponds in quantum mechanics to a particle in a Landau level with a
very large quantum number. If $ E_{\nu}$ be the energy of a Landau level
so that

$$     E_{\nu}=(\nu +\frac{1}{2})\hbar \Omega ,  \eqno {(1)}$$
where $\Omega=(eB/mc)$ is the gyrofrequency in the magnetic field B, then
$\nu \gg 1$ corresponds to the classical limit and $\nu \hbar=\mu$
defines the gyroaction, which classically has the form $ \mu=
\frac{1}{2}m{v_{\perp}^2}/\Omega$, ($v_{\perp}$ is the component of
velocity perpendicular to the magnetic field).

Consider now the propagation of an electron beam of a given energy
injected into the magnetic field with a small pitch angle $\delta$, so
that $v_{\perp}=v$ sin $\delta$, and $v_{\parallel}=v$ cos $\delta$,
$v_{\parallel}$ being the velocity parallel to the magnetic field. The
electrons in this beam are then in a group of Landau levels sharply peaked
around
the quantum number $\nu= E_{\perp}/{\hbar\Omega}=
\frac{1}{2}m{v_{\perp}}^2/{\hbar\Omega}$. For a typical laboratory
situation, if we choose E= 1 keV, and a magnetic field B= 100 g, then
$\nu \simeq {10}^8$, which is clearly $\gg 1$.

A charged particle in a magnetic field can be described by the Hamiltonian

$$ H = \frac{{P_{\parallel}}^2}{2m} + \frac{{P_\perp}^2}{2m} +\frac{1}{2}m
{\Omega}^2 {\xi}^2 ,
\  \eqno{(2)}$$
where $\xi$ represents the coordinate perpedicular to the magnetic
field and the potential energy term $\frac{1}{2}m{\Omega}^2 {\xi}^2$
represents the hamonic oscillator term corresponding to the Landau
gyro-oscillations, while the first term represents the free motion along
the magnetic field. As we shall see later, this Hamiltonian will also
describe, with some modification, the oscillatory motion of a diatomic
molecule.

Let $\chi_{\nu}$ represent the Landau eigenfunctions which are essentially
the harmonic oscillator wave fuctions [5]. Let there be a scatterer in the
path of the electron beam, a small obstacle, like the wires in grid
through which the electron beam may be made to pass. The scattering which
is assumed to
be elastic may kick the electrons from the Landau level $\nu$ to $\nu \pm
l$, where $\nu \gg l > 1$. If $\tilde H$ be the perturbation
Hamiltonian
which describes the scattering, then the transition amplitude for the
process is given by

$$ {\beta}^{(l)}_{\nu} \equiv \left< \nu-l \left|\tilde H\right| \nu
\right>
=\int d\xi \chi_{\nu-l}(\xi)\tilde H \chi_{\nu}(\xi) \, \eqno{(3)}$$

where $\xi$ is the coordinate normal to the magnetic field representing
the coordinate of the Landau gyro-oscillator.

Let $\phi_{\nu}$ represent the complete wave function of the particle in a
magnetic field including a plane wave corresponding to its free motion
along the magnetic field, so that 

$$ \phi_{\nu} = \chi_{\nu}(\xi)e^{i\kappa_{\nu}x}\   \eqno{(4)} $$

where

$$ \kappa_{\nu} = \frac{1}{\hbar}\left[2m(E-\nu \hbar
\Omega)\right]^{\frac{1}{2}}  \ \eqno{(5)}$$
and $x$ is the coordinate along the magnetic field, while $E$ is the total
energy of the particle. The transition amplitude including the
eigenfunction along the magnetic field is given by

$$ {\alpha}^{(l)}_{\nu} =\int d\xi {{\phi^{*}}_{\nu-l}} \tilde H
{\phi}_{\nu} 
                       =
{{\beta}^{(l)}_{\nu}} exp{\left[i\left(\kappa_{\nu}-\kappa_{\nu-l}\right)x
\right]}
\  \eqno{(6)}$$

Now making use of the assupmtion $l \ll \nu$, we expand ${\kappa}_{\nu-l}$
around ${\kappa}_{\nu}$ using the expression (5) which gives

$$ {\kappa}_{\nu}-{\kappa}_{\nu-l} \simeq l \frac{\partial
{\kappa}_{\nu}}{\partial \nu} =\frac{l \Omega}{v}\ \eqno{(7)}$$

where $v$ is the velocity of the particle along the magnetic field

$$ v= \left[\frac{2}{m}\left(E-\nu \hbar \Omega\right)\right]^{1/2}
\  \eqno{(8)}$$

The difference $({\kappa}_{\nu}-{\kappa}_{\nu-l})=l \Omega/v$ represents
the
change in the wave number ${\kappa}_{\nu}$ of the plane wave, in
consequence
of the change in the Landau level quantum number from $\nu$ to $\nu-l$ due
to the elastic scattering off the obstacle. The transition amplitude 
is thus given by 

 $$ {\alpha^{(l)}_{\nu}} =
{\beta^{(l)}_{\nu}} exp\left[i(l\Omega/v)x\right]
\ \eqno{(9)}$$

This transition amplitude is again a wave function, representing a plane
wave by virtue of the exponential factor. But as we notice, it is clearly
independent of $\hbar$. Since it is derived directly from the quantum
mechanical wave function (4), it represents a mattter wave with a wave
length $\lambda = 2\pi v/\Omega$. For an electron energy parallel to the
magnetic field $E_{\parallel}=500 eV$ and a magnetic field B= 100g, we
find $\lambda \simeq$ 5 cm. Thus this matter wave length falls in the
macrodomain in contrast to the usual de Broglie wave length which is
generally in the {\AA}  range.

It may be mentioned that these ideas have been more formally expressed in
a recent paper by the author[2] where he has derived a set of Schr${\rm
\ddot o}$dinger-like equations starting from the quantum mechanic
Schr${\rm
\ddot o}$dinger equation (in its path integral representation) for the
charged particle dynamics in a magnetic field. These are

$$ \frac{i\mu}{l}\frac{\partial \Psi(l)}{\partial t} =
-\biggl(\frac{\mu}{l}\biggr)^2 \frac{\partial^2 \Psi(l)}{\partial x^2}
+\bigl(\mu
\Omega \bigr)\Psi(l),\ \ l=1,2,..\  \eqno{(10)}$$
where $\mu$, which is the gyroaction and has been shown in typical
laboratory conditions to be $\sim 10^8\hbar$, is a classical object and
appears in the role of $\hbar$ in these equations. In terms of the
notation of the foregoing treatment $\mu =\nu \hbar$ with $\nu \gg 1$.
Furthermore, the wave functions $\Psi(l)$ of these equations are actually
the transition amplitudes as defined above, for the quantum mechanical
state with a large Landau quantum number $\nu$ to the one with
quantum number $\nu-l$, induced by a perturbation. The number $l$ labels
this wave function as
$\Psi(l)$.

By virtue of the large (classical) value of $\mu$ which appears in the
place of  $\hbar$, these equations for the amplitude functions $\Psi(l)$
describe matter wave phenomena in the macrodomain of classical mechanics.
This is essentially equivalent to what has been demonstrated above in
Eq.(9), in more direct manner.

The wave function of Eq.(9), as also the Schr${\rm \ddot o}$dinger-like
equations, predict the matter wave phenomena with the wave length of a few
centimeters for the charged particles moving along a magnetic field. The
one-dimensional matter wave interference phenomena which correspond
to these macroscopic wave functions of the form (9) have indeed been
observed by the author and his coworkers [3,4].

The experimental results reported in Ref. [3,4] exhibit the existence of
discrete energy bands (the maxima and minima) in the transmission of
electrons along a magnetic field, when the energy from a gun is swept as
they transit from the latter along a magnetic field to a detector plate a
distance
$L_p$ away. These bands which are rather unexpected in the
parameter domain of the experiments where classical mechanical equations
of motion are supposed to operate, have been identified as the
interference maxima and minima in the energy domain, with a (nonquantal)
macroscopic wave length $\lambda = 2 \pi v_{\parallel}/\Omega$, in
accordance with the form (9) of the macroscopic wave function (where
$v_{\parallel}$ is the electron velocity along the magnetic field and
$\Omega = eB/mc$, the electron-gyrofrequency in the magnetic field $B$).
The interpeak separation of the transmission bands (in energy) are
found to be inversely proportional to the distance $L_p$, so that the
latter corresponds to a frequency as the energy is swept.  
The experiments of Ref.[3,4] thus confirm the predictions of the theory
on the existence of the macroscopic form of the matter waves. 

We have also
found the existence of beat phenomena [6] in these experiments - a
modulating beat structure of the already reported discrete energy band
structure when the two ``frequencies'' in the system are close together.
In the presence of a grounded grid at a distance $L_g$ from the electron
gun,
one has two frequencies in the system corresponding to the $L_p$ and $L_g$
and the frequency of the observed beats in the transmitted signal is found
to correspond to the difference $(L_p-L_g)$, where $(L_p-L_g) \ll
L_p$. This is just what occurs in other wave phenomena as well. These
observed
beats
thus constitute a further, even tighter evidence for the wave behaviour
of particles moving along a magnetic field.

{\bf 3. Macroscopic matter waves for composite systems in their high
internal state of excitation.}

Having discussed the concept of macroscopic matter waves for charged
particles in a magnetic field in the last section, whose wave
manifestations in the macrodomain have also been observed, we now extend
these considerations to other composite systems such as atoms and
molecules in their internal state of excitation . We first discuss some
gedanken experiments to point out the possible macroscopic wave
manifestations of these systems and then discuss the possibility of
carrying out real experiments to observe these manifestations associated
with such composite systems.

{\bf A. Diatomic molecule in a highly excited vibrational state}

First we consider a diatomic molecule in a highly excited vibrational
state ignoring for the moment its rotational and electronic degrees of
freedom. Such a system is described by the same Hamiltonian as (2) except
for changing the 'parallel' momentum $p_{\parallel}$ to the 
momentum ${\bf P}$ of the centre of mass $M$ of the diatomic molecule and
changing the
'perpendicular' momentum $p_{\perp}$ to the momentum ${\bf p}$ of the
reduces mass $m$ , and identifying $\xi$ as the reduced mass coordinate,
so that we have the Hamiltonian as

$$ H^{v}_{DA}= \frac{P^2}{2M}+ \frac{p^2}{2m}+\frac{1}{2}m{\omega}^2
{\xi}^2 
\, \eqno{(11)}$$
where we have now the vibrational frequency $\omega$ of the diatomic
molecule , and we have of course ignored
the anharmonic terms for simplicity.

If we employ a similar notation as before then the eigenfunction for the
system with the Hamiltonian ${H^{v}_{DA}}$ corresponding to
the free
motion of the centre of mass with momentum ${\bf P}$, and vibrational
state $\nu$ is given by

$$\psi ({\bf P}, \nu ) = A_1\ e^{i {\bf P} \cdot {\bf X}/\hbar} {
\chi}_{\nu} (\xi)
\ , \eqno{(12)} $$
with the total energy $E$ given by
$$E = \frac{P^2}{2M} + \hbar \omega \left(\nu + \frac{1}{2} \right) \ ,
\eqno{(13)}$$
where $\chi_{\nu} (\xi)$ are the normalized harmonic oscillator wave
functions.

Consider now a beam of such particles with a given momentum ${\bf P}$ and
in a highly excited vibrational state $\nu \gg 1$, which can be prepared
using appropriate laser techniques. Let the beam be scattered by a grid of
scatterers $G_1$ in its path, with small transverse dimensions. Assume
that the scattering is elastic with respect to the total energy $E$ of the
particle, and the scattering changes only its internal vibrational state
to $\nu^{\prime}$. So, the final state after the scattering is 
$$\Psi^{\prime} = A^{\prime}_1 e^{i {\bf P}^{\prime}\cdot{\bf X}/\hbar}
\chi_{\nu^{\prime}} (\xi)\,  \eqno{(14)}$$ 
where $\mid \nu^{\prime} - \nu
\mid \ll \nu$. If then $\tilde{H}(\xi)$ is the perturbing Hamiltonian
which causes the scattering, then the transition amplitude is given by %
$$\alpha_{\nu^{\prime}\nu} = \left< \nu^{\prime} \mid \tilde{H} \mid \nu
\right> = A_1 \ A^{\prime}_1 \exp \left[ - i \left( {\bf P}^{\prime} -
{\bf P}\right) \cdot {\bf X}/\hbar \right] \int d \xi \chi_{\nu^{\prime}}
(\xi) \tilde{H} (\xi) \chi_{\nu} (\xi)\, \eqno{(15)} $$ where $A_1$ and
$A^{\prime}_1 $ are appropriate normalization constants.

The change $\Delta E_i$ in the internal energy $E_i$ is
$\Delta E_i = \hbar \omega \left( \nu^{\prime} - \nu \right) = \hbar \omega l
 $,
where $l \ll \nu$. Taking ${\bf P}$ and ${\bf P}^{\prime}$ to be
predominantly
along a given X-direction, or alternatively, choosing for simplicity ${\bf
P}^{\prime}$ to be in the direction of {\bf P} (one can do so
experimentally) we have, because of the elastic nature of the scattering,
the  expressions:
$ P = \left[ 2M \left( E - \nu \hbar \omega \right) \right]^{1/2}$ and
$P^{\prime} = \left[ 2M \left( E - \nu^{\prime} \hbar \omega \right)
\right]^{1/2} $

On expanding $P^{\prime}$ around $P$ for $l \ll \nu$, one gets to lowest order
in $l\hbar$ :
$P^{\prime} - P \approx  \left[ 2 \left( E - \nu \hbar \omega \right)/M
\right]^{-1/2} l \hbar \omega $.
Using this in (15) yields the transition amplitude as
$$\alpha^{(1)}_{\nu^{\prime}\nu} (X) = A_1 A_1^{\prime} \beta_{\nu^{\prime}\nu} \exp
\left[ i l \frac{\omega (X - X_1)}{v} \right]\, \eqno{(16)}$$
where $\beta_{\nu^{\prime}\nu}$ is the matrix element of the perturbation
$\tilde{H}(\xi)$ between the oscillator states $\nu^{\prime}$ and $\nu$ and $v$
is the magnitude of beam velocity $v = \left[2 \left(E - \nu \hbar \omega
\right) \right/M ]^{1/2}$. Note that while the expression for $v$ does
involve $\hbar$, but in the limit $\nu \gg 1$, $\nu \hbar$ is actually a
classical object $I=\nu \hbar, (\nu \gg 1)$. Also from the experimental
point of view $v$ is simply the beam speed of the particles and it is not
relevant what its expression is in terms the total energy $E$ and the
internal state energy $E_i=I\omega$. Similar remarks apply to the other
cases as well
discussed later. 

The form of the perturbation $\tilde{H}$ is
left general enough, and may be given any specific form as required. The
important thing to note is that the transition amplitude for the
translational centre of mass degree of freedom has the form $\exp \left[ i l
\frac{\omega X}{v} \right]$ and is independent of $\hbar$. Note
that in general the scattering at the grid $G_1$ could lead to different
values of $l = 1, 2, 3 \cdots$ . But we keep only any one value for the
simplicity of discussion. 

Consider next another grid $G_2$ of scatterers located at the point $X_2$ in the
path of the beam, the scattering (transition) amplitude from this grid is
given by
$$\alpha_{\nu^{\prime}\nu}^{(2)} (X) = A_2 A_2^{\prime} \beta_{\nu^{\prime}\nu} \exp
\left[ \frac{i l \omega}{v} \left( X - X_2 \right) \right] \eqno{(17)}$$
where $A_2$  and $A^{\prime}_2$ are again appropriate normalization
constants.

Note that the expressions (16) and (17) represent wave functions
corresponding to a wave number
$k_l = \frac{l \omega}{v} = l k $
which is the $l^{th}$ harmonic of the basic wave number $k = \omega/v$, the
corresponding wave length being $\lambda = 2 \pi v/\omega$. This is clearly
independent of $\hbar$, and could lie in the macro or meso-domain. For a
typical
diatomic molecule, the vibrational wave number is $\left( \omega/2 \pi
c\right)
\approx 2 \times 10^3 cm^{-1}$. Taking a modest value of beam velocity $v
\approx 10^8 cm^{-1}$, this gives $\lambda \approx 0.1 \mu$. This is about
three orders of magnitude larger than the typical de Broglie wave length of a
few $\stackrel{\rm o}{\rm A}$.

One can now look for interference between the waves given by (16) and (17)
originating at the scatterer grids at $X_1$ and $X_2$. At a point $X$
downstream of the grids at $X_1$ and $X_2$ the total amplitude is given by
$$\alpha = \alpha^{(1)}_{\nu^{\prime}\nu} + \alpha^{(2)}_{\nu^{\prime}\nu} = e^{i l k
X} \beta_{\nu^{\prime}\nu} \left[ A_1 A_1^{\prime} e^{-ilkX_1} + A_2 A_2^{\prime}
e^{-ilkX_2} \right] \eqno{(18)}$$
whence the intensity of the scattered particles is given by
$$\mid \alpha \mid^2 =\mid  \beta_{\nu^{\prime}\nu}\mid^2 \left\{
\left(A_1 A_1^{\prime}
\right)^2 + \left( A_2 A_2^{\prime} \right)^2 + 2 \left( A_1 A_1^{\prime}
A_2
A_2^{\prime}\right) \cos \left[ l k \left( X_1 - X_2 \right) \right] \right\}
\eqno{(19)}$$ 

This therefore describes interference maxima and minima through the $\cos
\left[ k l \left( X_1 - X_2 \right) \right]$ term which is independent of
$\hbar$ and hence belongs to a nonquantal domain. Such interference effects
should be present in the experimental arrangement described above. This is
analogous to the double slit interference, the grids at $X_1$ and $X_2$
corresponding to the two slits, but now in one dimension. One can check the
validity of the expression (19), by working with different diatomic
molecule
to vary $\omega$, and different beam velocities to check the dependence on
$v$, as well as different values of $\left( X_1 - X_2\right)$. 

Though we have considered only one value of $l$, there would, in general
exist many values of $l=1, 2, 3...$ in the excitation spectrum of the
vibrational states of the diatomic molecule as a result of scattering
off the grids $G_1$ and $G_2$. However, $l=1$ would be the most dominant
one, being closet in energy to the central quantum number $\nu$. There
would thus exist interference maxima, most dominant ones for $l=1$ and
successively subdominant ones for $l=2, 3, ...$. On the other hand, the
higher harmonics may lead to appropriate change of shape of the peaks
corresponding to the fundamental.

There would arise, however, other more practical difficulties, if one were
to move from the plane of the gedanken experiments to  more real
ones. One of them would be the preparation of the vibrational states with
quantum numbers strongly peaked around a $\nu, (\nu \gg 1)$. This we
assume can be carried out with appropriate laser techniques. There may be
other technical difficulties, as for instance, the preparation of a beam
of such excited molecules with a given well defined velocity. While there
will certainly be such experimental difficulties, our main objective here
is to point out the existence, in principle, of the macroscopic and
mesoscopic matter waves rather than to address all the practical
difficulties, which we hope can be overcome with some experimental
ingenuity.

{\bf B. Charged particles in a magnetic field}

If we apply these considerations to the case of charged particle in a
magnetic field, we have the following correspondence
$\omega \Rightarrow \Omega = eB/mc$, and
$v \Rightarrow v_{\parallel} = \left[ 2 \left( {\cal E} - \mu \Omega
\right)/m \right]^{1/2} $

In the case of the charged particle in a magnetic field, the bound 
vibrational state of the diatomic corresponds to the Landau state ( bound
to the magnetic field in the perpendicular direction) and the
free motion of the centre of mass corresponds to the free motion of the
charged particle along the magnetic field. 
The wave number $k$ now has the form
$k = \Omega/v_{\parallel}$
with the wave length $\lambda = 2 \pi v_{\parallel}/\Omega$. This is
precisely the wave length which follows from the form (9) of the
macroscopic wave function and Eq.(10) of the formalism[1,2]
corresponding to the mode $l=1$. The  discrete
energy bands observed in the experimental results reported [3,4] as well
as the beat structure observed more recently [6] are indeed a
manifestation of matter waves in macro-dimension, for $l=1$. In fact the
experimental
results have indeed exhibited the dependence of the
locations of the interference maxima quite decisively on the parameters
$\left( X_1 - X_2 \right)$, and $\Omega = eB/mc$ and $v$ through $k =
\Omega/v$ in the manner required by the expression (19). Note that the
mode number $l$ in Eq.(1) corresponds precisely to the harmonic number $l$
in Eq.(9) and (19). This thus vindicates
the formalism of Ref.[1,2] whose predictions motivated these experiments
in
the first place. 

It may be noted that for a
magnetic field of around 150 g and $v_{\parallel} = 10^9 cm s^{-1}$, the wave
length $\lambda = 2 \pi v_{\parallel}/\Omega \approx 2.6\ cm$. This is
clearly of macroscopic dimensions.\\\\ 
{\bf C. Rotational and vibrational states of a diatomic molecule}

If we next think of the case of rotational states of a diatomic molecule, they
do not obviously belong to the generic Hamiltonian (11). Yet with some
difference such a system can also be shown to exhibit matter waves in
macro-dimensions through wave amplitudes of the form (16) and (17). We
recall
that the origin of the form of the amplitude (16) is that the quantity
$\left( P -  P^{\prime}\right)/\hbar$ becomes independent of
$\hbar$ to lowest order, when $P$ and $P^{\prime}$ are substituted for. For the
case of rotational states, $E_j = K \hbar^2 j (j+1) \simeq K \hbar^2 j^2$
for highly
excited states. Then $P = \left[ 2 M \left( E - K \hbar^2 j^2 \right)
\right]^{1/2}$ and $P^{\prime} = \left[ 2M \left(E - K \hbar^2
j^{\prime^2}\right) \right]^{1/2}$. This gives
$P - P^{\prime} \simeq 2 K \hbar^2 j l \left[ 2 \left( E - K \hbar^2 j^2
\right)/M \right]^{-1/2} $

If we now define $J = \hbar j (j \gg 1)$, as the angular momentum value in the
large quantum number limit, we get
$ \left( P - P^{\prime} \right)/\hbar \simeq \left(2 J K/v \right) l $
where $\left( j^{\prime} - j = l \ll j \right)$ and where

$v = \left[ 2 \left( E - K \hbar^2 j^2 \right)/M \right]^{1/2}$. It
may
be remarked that while $J=\hbar j$ and likewise $v$ may not
appear to be independent of $\hbar$, they could be considered to be
effectively so in the large quantum number limit $j\gg 1$, as in this
limit
$J=\hbar j$ could be considered a classical angular momentum. We
thus see that looked at this way, $\left( P - P^{\prime}\right)/\hbar$ may
be taken to be     
independent of $\hbar$, and the wave amplitude $\alpha_{j^{\prime} j}$
takes the form
$$   \alpha_{j^{\prime} j} = A_1 A_1^{\prime} \gamma_{j^{\prime} j} \exp
\left[ i l \frac{2 KJ}{v} \left( X - X_1\right) \right] $$
$$               = A_1 A_1^{\prime} \gamma_{j^{\prime}j} \exp \left[
i
\frac{l \omega_j}{v} \left( X - X_1 \right) \right] \eqno{(20)}$$
where $KJ = J(2 m R^2)^{-1}$, $m$ being the reduced mass, and $R$ the
internuclear distance. $2KJ$ is then $J/mR^2$ and is of the dimension of
angular velocity. Thus we have  written $\omega_j = J/mR^2$ in the second
line of the above equation. The
amplitude
$\alpha_{j^{\prime}j}$ is then
again of the form (16), where $\gamma_{j^{\prime}j}$ has a meaning
corresponding to that  for $\beta_{\nu^{\prime}\nu}$ in (16). The
amplitude
(20) is again $\hbar$ independent and corresponds to a wave length of meso
or
macrodimension. However, the important difference between this case and the
vibrational case is that while $\omega$ in the latter case is independent of
the quantum state, $\omega_j$ increases linearly with the quantum number $j$.

In general, however, the scattering by the grid of scatterers would induce
transitions in vibrational as well as in rotational states. One would then
have a general wave amplitude of the form :
$$\alpha_{j^{\prime}j\; ; \; \nu^{\prime}\nu} = A_1 A_1^{\prime} \Gamma_{j^{\prime}j
\; \nu^{\prime}\nu} \exp \left[ \frac{i}{v} \left( l_{\nu} \omega_{\nu} +
l^{\prime}_j \omega_j \right) \left( X - X_1 \right) \right] \eqno{(21)}$$
where now we mean by $\omega_{\nu}$, the vibrational frequency and by
$\omega_j$, the
rotational frequency. We define the corresponding wave numbers as $k_{\nu} =
\omega_{\nu}/v$ and $k_j = \omega_j /v$ where $v$ is now 
$v = \left[ 2 \left( E - \nu \hbar \omega_{\nu} - K j^2 \hbar^2 \right)/M 
\right]^{1/2}$\\\\ 

We thus have a more general expression than (19) for this case involving
both the vibrational and rotational excitations

$$ \mid \alpha \mid^2 =\mid \Gamma_{j^{\prime}j \nu^{\prime} \nu}\mid^2
\Bigl\{(A_1 A^{\prime}_2)^2 +A_2 A^{\prime}_2)^2 +2(A_1 A^{\prime}_1 A_2
A^{\prime}_2) cos\left[(l_{\nu} k_{\nu}+ l_j k_j)(X_1-X_2)\right]\Bigr\}
, \   \eqno{(22)}$$

As we know, the vibrational frequency of a diatomic molecule $\omega_{\nu}
\gg \omega_j $, the rotational frequency. Therefore, analogously to the
case with
the vibrational-rotational spectrum of the diatomic molecules, the
spacing
of the fringes corresponding to the rotational matter wave number $ k_j=
{\omega_\nu}/v$ would be corespondingly greater compared to the those
relating to the vibrational matter wave number
$k_{\nu}={\omega_\nu}/v$. Here the vibrational fringes will be contained
within the rotational fringes, contrary to the case of the
vibrational-rotational spectum. It would thus be possible to differentiate
the two sets of fringes with respect to the variation of the spatial
separation $(X_1-X_2)$.

{\bf D. Rydberg states of an atom}

One may also consider the Rydberg states of an atom for a similar
discussion, which
presents a rather interesting case. The energy levels for this case are given
by
$$E_n = - \frac{me^4}{2 \hbar^2 n^2} \ .$$
Using the expressions $P = \left[ 2 M \left(E - E_n \right) \right]^{1/2}$
and $P^{\prime} = \left[2M \left( E - E_{n^{\prime}}\right) \right]^{1/2}$, we
find
$P - P^{\prime} \approx \frac{2 l E_n}{n v}$
where $v = \left[ 2 \left( E - E_n \right)/M \right]^{1/2}$ is again the
magnitude of the centre of mass velocity in the highly excited state $n$,
with the total energy being $E$, and where $l = \mid n^{\prime} - n \mid \ll
n$ is assumed. Using the foregoing expression for $P^{\prime}-P$ in the
expression corresponding to
(16) gives
$$\alpha_{n^{\prime}n} = A_1 A_1^{\prime}\;\Delta_{n^{\prime}n} \exp \left[ i
\frac{2 l E_n}{nv \hbar} \left( X - X_1 \right) \right] \ . \eqno{(23)}$$
where $\Delta_{n^{\prime}n}$ is the scattering (transition) amplitude for a
perturbation $\tilde{H}$ between the Rydberg states $n^{\prime}$ and $n$. The
expression (23), of course, involves $\hbar$ unlike the expressions
(16) and
(17) for the harmonic internal degree of freedom, which are independent of
$\hbar$. But because of the presence of $n$ in the denominator of the
exponent of (23), the effective wave number
$$k_n = \frac{2E_n}{v (n \hbar)} = \frac{\omega_n}{v} \ ; \ \omega_n =
\frac{2 E_n}{n \hbar} \ ,$$
can be small enough to be of macroscopic or mesoscopic dimensions if $n$
is taken to be
sufficiently large. For $n = 100, \ \omega_n \approx 6.6 \times 10^{10}
\;{\rm rad}\;s^{-1}$, and $k_{\parallel} \approx 660\;cm^{-1}$ for $v =
10^8\;cm\;s^{-1}$. This gives $\lambda_n = 2\pi/k_n \approx
{10}^{-2}\;cm$, which
is of macroscopic, or if one prefers, of mesoscopic dimensions.\\\\
{\bf Discussion and summary}

The concept of macroscopic and mesoscopic matter waves as distinct from
the deBroglie matter waves that has been presented above, first arose in
relation to the dynamics of charged particles in a magnetic field as
described by a probability amplitude theory [represented by a set of
Schr${\ddot o}$dinger- like equations] given by the author[1] in
1985. This was obtained as a Hilbert space representation of the classical
Liouville equation for the system, and by virtue of its amplitude
character, it predicted the existence of matter wave phenomena with a
macroscopic wave length, namely the one-dimensional interference effects.
The role of $\hbar$ in these equations is enacted by a macroscopic action
$\mu =\nu \hbar, (\nu \gg 1)$.

Recently the author has been able to derive [2] the same set of
Schr${\ddot o}$dinger-like equations starting from the quantum
mechanic Schr${\ddot o}$dinger equation. It is the macroscopic
magnitude of $\mu$ ( which replaces $\hbar$) that is responsible for the
macroscopic matter waves in charged particle dynamics along a magnetic
field. Indeed the one-dimensional interference effects with macroscopic
matter waves with wave length typically in the range $\sim 5 cm$ have been
reported by the author and his collaborators [3,4,6]

A close examination of the derivation of Ref.[2] gave the author the clue
that such macroscopic and mesoscopic matter waves should be possible with,
in fact, any composite system, such as atoms and molecules in their highly
excited internal states. The discussion presented in this paper is an
exposition these concepts as applied to diatomic molecules and Rydberg
atoms as illustrative examples.

As already mentioned, the interesting point about these matter waves is
that their wave length is not related to the mass of the particle as with
the deBroglie waves, but rather to the frequency $\omega$  associated
with the highly excited state of its internal degree of freedom-
vibrational, rotational for a molecule, and Landau gyro-oscillations for
charged particles in a magnetic field, or the frequency associated with
atomic levels in an atom. 
and is indepedent of
the Planck quantum in this correspondence limit. In a sense these waves
may be regarded as classical counterpart of the quantum deBroglie waves.

It is also clear from the above discussion that such kind of meso or
macrodimension matter waves are generically possible. It would be interesting
to try to observe them with  atoms and diatomic molecules or even
nuclei, just as we have observed them in the system of charged particles in a
magnetic field [3,4,6]

The existence of these waves has already been experimentally
demonstrated [3,4,6]
for electrons propagating along a magnetic field, following initially its 
predictions by a quantum like theory  given by the author[1,2] 
which actually stimulated the entire line of investigation culminating in
the findings presented here. We now predict the existence of such waves
for other systems, atoms and molecules, or any other composite systems
which could be looked for experimentally.

Acknowledgments: The author would like to thank Prof. J.C. Parikh and
Rajat Varma for going through the manuscript and making suggestions for
improvement.

\newpage
\noindent
{\bf References}
\begin{enumerate}
\item R.K.Varma, Phys. Rev. {\bf A31} 3951-3959 (1985). 
\item R.K.Varma, Phys. Rev. E{\bf 64} 036608 (2001).
\item R.K.Varma and . Punithavelu, Mod. Phys. Lett. {\bf A8} 167-176
(1993). 
\item R.K.Varma and A.M.Punithavelu, A.M. {\it ibid.} {\bf A8} 3823-3834 (1993).
\item L.D.Landau and E. M. Lifshitz, {\it Quantum Mechanics} Addison-
Weseley Publishing Co. Reading, Mass. 1965, p.425
\item R.K.Varma, A.M.Punithavelu and S.B.Banerjee, {\it Observation of
beat
structure in the transmission of electrons along a magnetic field in the
classical mechanical domain}, Phys. Rev. E {\bf 65} Feb. (2002)
\end{enumerate}
\end{document}